\documentclass[twocolumn,aps, preprintnumbers,superscriptaddress]{revtex4}
\usepackage{graphicx}
\usepackage{graphicx}
\usepackage{dcolumn}
\usepackage{bm}
\usepackage{amsmath, amsthm, amssymb}

\usepackage{color}

\begin{document}
\title{Revisit Kerker's conditions by means of the  phase diagram}
\author{Jeng Yi Lee}
\affiliation{
Institute of Photonics Technologies, National Tsing Hua University, Hsinchu 300, Taiwan}

\author{Andrey E. Miroshnichenko}
\affiliation{Nonlinear Physics Centre, Research School of Physics and Engineering, The Australian National University, Canberra 2601, Australia}

\author{Ray-Kuang Lee$^{*}$}
\affiliation{
Institute of Photonics Technologies, National Tsing Hua University, Hsinchu 300, Taiwan}
\affiliation{Physics Division, National Center for Theoretical Sciences, Hsinchu 300, Taiwan}
\date{\today}
\affiliation{Corresponding author: rklee@ee.nthu.edu.tw}

\begin{abstract}
For passive electromagnetic scatterers, we explore a variety of extreme limits on directional scattering patterns in phase diagram, regardless of details on the geometric configurations and material properties.
By demonstrating the extinction cross-sections with the power conservation intrinsically embedded in phase diagram,  we give an alternative interpretation for Kerker first and second conditions, associated with zero backward scattering (ZBS) and nearly zero forward scattering (NZFS).
The physical boundary and limitation for these directional radiations are illustrated, along with a generalized Kerker condition with implicit parameters.
By taking the dispersion relations of gold-silicon  core-shell nanoparticles into account, based on the of phase diagram, we reveal the realistic parameters to experimentally implement ZBS and NZFS at optical frequencies.
\end{abstract}
%\pacs{XXXX}

\maketitle

In 1983, Milton Kerker \textit{et al.} revealed that under a proper combination of magnetic and electric dipoles of a subwavelength magneto-dielectric particle,  one can have  an asymmetric field radiation with zero backward scattering (ZBS) or zero forward scattering (ZFS), which are known as  first and second Kerker's conditions, respectively ~\cite{kerker}. 
For ZBS, the permeability $\mu$ and permittivity $\epsilon$ must have the same value, i.e., $\mu=\epsilon$; while for ZFS, the  permittivity and permeability need to satisfy the condition: $\epsilon=(4-\mu)/(2\mu+1)$,  in the quasistatic limit.
Such an asymmetric radiation in backward or forward directions is also associated with a directional Fano resonance, due to an induced resonant  wave forming the constructive or destructive interferences to background scattered one  ~\cite{fano,fano1}.

Experimental realizations toward directional scattering patterns have been demonstrated with a single nanoparticle made of GaAs ~\cite{backwardexp}, silicon ~\cite{forwardexp} or gold nano-antenna~\cite{general}. 
Moreover, due to the lack of a giant magnetization in natural materials at optical wavelengths, several ways are proposed to realize anomalous asymmetric radiations, such as inducing interferences among dipole and quadrupole channels on metallic-dielectric nanoparticles~\cite{new1},  using gain dielectric nanoparticles to compensate inherently scattering loss in order to have zero extinction cross-section~\cite{new2}, using nanoparticles made of high index of refraction optical materials ~\cite{magnetic, Science2016}, and engineering  aluminum nanostructures in shape of pyramid to induce a magnetic dipole~\cite{new3}.

However, when the optical theorem is taken into consideration, an inconsistency arises because the electric scattering amplitude in forward direction is related to the total extinction cross-section~\cite{opticaltheorem}. 
That is, the ZFS implies zero extinction cross-section. 
The consequence of zero extinction strictly requires the zeros of absorption and scattering cross-sections.
As a result, ZFS is an impossible phenomenon for any non-invisible passive systems.
Although  many studies support the trends of such asymmetrical radiation ~\cite{exp3,exp2,break,nanodisk,acs1,acs2}, a study on the limitation and physical boundary regardless of scattering system details is still lack.

Recently, based on energy conservation law, i.e., the total amount of outgoing electromagnetic power by passive objects should be the same or smaller than that of incoming one, we propose a phase diagram to reveal all allowable scattering solutions and energy competitions among absorption and scattering powers~\cite{phase}.
In this work,  we take the advantage of this phase diagram for passive electromagnetic scatterers to revisit first and second Kerker's conditions,  in order to reveal the limits on directional scattering patterns, regardless of details on the geometric configurations and material properties.
For ZFS, we find that forming a perfect destructive interference for a pair of scattering coefficients by electric and magnetic dipoles is forbidden for any non-invisible system, resulting in the support of {\it nearly} zero forward scattering (NZFS) only.
Any absorption loss will destroy the required out-of-phase conditions, leading to  a non-vanished extinction cross-section  intrinsically.
Nevertheless,  for ZBS, there is no such constraint from the  extinction cross-section, as long as the corresponding scattering coefficients are the same within the  allowable regions in  phase diagram.

\begin{figure*}[ht]
\begin{center}
\includegraphics[width=18.0cm]{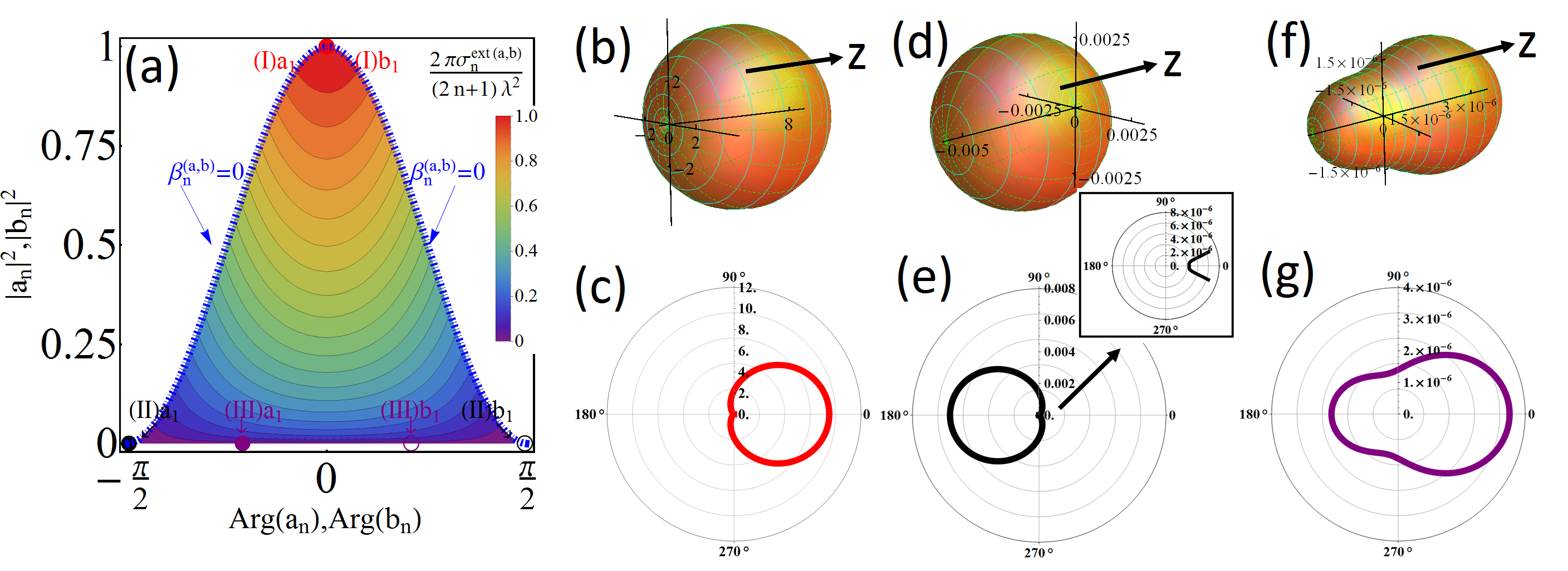}
\end{center}
\caption{(a) Phase diagram for the extinction cross-section, defined by amplitude square ($|a_n|^2$, $|b_n|^2$) and phase ($\text{Arg}(a_n)$, $\text{Arg}(b_n)$) of scattering coefficients for passive scatterers. 
The colored region indicates the allowed solutions, with the contour line represents a constant normalized extinction cross-section ($2\pi \sigma_n^{\text{ext} (a,b)}/(2n+1)\lambda^2$) in  TM or TE modes.
Three sets of supported scattering states are marked as (I), (II), and (III), with the corresponding 
3D-2D radiation patterns depicted in  (b-c), (d-e), and (f-g), respectively. 
Here, (b-c) reveals the ZBS condition; (d-e) reveals the nearly ZFS condition, as a residue of  scattering contributions  can still be found in the enlarged window in forward direction (see the Inset); (f-g) reveals a state with the same extinction cross-section as that of (d-e) (note the scale), but with losses.
The size of  scatterer used in all simulations is the same by setting $a=10^{-2}\lambda$.}
\end{figure*}

Moreover, we provide a set of permeability and permittivity with an additional intrinsic degree of freedom to support NZFS by employing the inverse method developed from the phase digram to design functional devices with required optical response ~\cite{phase}.
In particular, we consider core-shell nanostructures,  made of gold in core and  silicon in shell, as a target to realize ZBS and NZFS  at optical wavelength from $450$nm to $800$nm.
By taking into the real material dispersion relation into account, we reveal the working wavelengths to  demonstrate ZBS and NZFS with the same geometric configuration.
With the help of phase diagram, our results not only provide a deeper understanding on  light  scattering patterns, but also offer as a universal way  to design functional scatterers for light-harvesting, metasurface, nano-antenna, and nano-sensor.

We start with light illuminating on an isolated spherical object, by considering a  linearly $\hat{x}$-polarized electromagnetic plane wave propagating along the $z$-direction with time evolution $e^{-i\omega t}$.
By applying the spherical multiple decomposition, the scattered field can be represented with two sets of scattering coefficients $a_{n}$ and $b_{n}$, which correspond to the transverse magnetic (TM) and transverse electric (TE) modes, respectively.
Here, the index $n$ is used to denote $n$-th order spherical harmonic channel, and the corresponding absorption, scattering, and extinction cross-sections can be derived as follows ~\cite{phase,book1}:
\begin{eqnarray}
\sigma^{abs}&=&\sum_{n=1}^{n=\infty}\frac{(2n+1)\lambda^{2}}{2\pi}(\text{Re}\{a_{n}\}-\vert a_{n}\vert^2+\text{Re}\{b_{n}\}-\vert b_{n}\vert^2),\nonumber\\
\sigma^{scat}&=&\sum_{n=1}^{n=\infty}\frac{(2n+1)\lambda^{2}}{2\pi}(\vert a_{n}\vert^2+\vert b_{n}\vert^2),\nonumber\\
\sigma^{ext}&=&\sigma^{abs}+\sigma^{scat},\nonumber
\end{eqnarray}
with the  wavelength of incident plane wave $\lambda$. 
As a rule of thumb, the main contribution to these convergent series would be related to the size parameter, defined as $k_0\,a$, where $a$ denotes the radius of this spherical object and $k_0=2\pi/\lambda$ is environmental wavenumber.

By Kirchhoff vector integral and asymptotic analysis, the extinction cross-section can be expressed in terms of   the scattered electric field in the forward direction $\theta=0$ ~\cite{jackson}:
\begin{eqnarray} 
\sigma^{ext}=\frac{4\pi}{k_{0}}Im[\hat{x}\cdot\vec{f}(\theta=0)],
\end{eqnarray}
here, in the radiation regime (the observer is far from the scatterers) the scattered electric field can be approximated as $\vec{E}_{s}\rightarrow  E_{0}\vec{f}(\theta,\phi)e^{ik_{0}r}/r$  with the the strength of incident electric field  $E_{0}$.
It is also known that Eq. (1) corresponds to the optical theorem, which implies that any loss, no matter from intrinsic material loss or scattering radiation, will lead to the extinction, equivalently to the enhancement of scattered field in the forward direction.

Due to the inherently non-negative values for $\sigma^{abs}$ and $\sigma^{scat}$ in passive scatterers, it is obvious that  a perfect zero extinction cross-section exists only when both the absorption and scattering cross-sections vanish.
A zero extinction cross-section represents zero scattering electric field in the forward direction, implies that the nanopartcle produces no shadow.
This simple argument suggests that  for any passive scattering system only invisible bodies have no shadow, but one can embed active materials to compensate scattering loss in order to achieve a zero extinction ~\cite{new2}.

To study directional scattering patterns,  in terms of spherical harmonic orders  we can calculate the differential scattering cross-sections in the forward ($\theta=0$) and backward ($\theta=\pi$) directions, i.e., defined as $\sigma^{fw}$ and  $\sigma^{bw}$, respectively~\cite{opticaltheorem, book1}:
\begin{eqnarray}
\sigma^{fw}&=&\frac{\lambda^{2}}{16\pi}\vert \sum_{n=1}^{\infty}(2n+1)(a_{n}+b_{n})\vert^{2},\\
\sigma^{bw}&=&\frac{\lambda^{2}}{16\pi}\vert \sum_{n=1}^{\infty}(-1)^{n}(2n+1)(b_{n}-a_{n})\vert^{2}.
\end{eqnarray}
Here, by following the original concept from Kerker \textit{et al.}~\cite{kerker}  to eliminate the forward or backward scattered radiations, we have ZFS  or ZBS when $\sigma^{fw}=0$ or  $\sigma^{bw}=0$ is satisfied, respectively.
Conducted from Eq. (2), clearly, one can see that the key point for cancellation in the forward scattering ($\sigma^{fw}=0$ ) relies on the destructive interferences between  $a_n$ and $b_n$, i.e., $a_n = - b_n$, corresponding to the same magnitude and  out-of-phase condition in the two scattering coefficients.
As for  ZBS ($\sigma^{bw}=0$), we need to have two equal scattering coefficients in the same spherical harmonic orders, i.e., $a_n = b_n$.

Now, we turn to the phase diagram for passive electromagnetic scatterers, by applying the {\it phasor} representation to the scattering coefficients, i.e., defining $ a_{n}=\vert a_{n}\vert e^{i \text{Arg}(a_{n})}$ and $ b_{n}=\vert b_{n}\vert e^{i \text{Arg}(b_{n})}$. 
As we reported in Ref. ~\cite{phase},  the power conservation intrinsically holds for the absorption cross-sections, $\sigma_{n}^{abs(a,b)}\geq 0$ in each spherical harmonic order, which corresponds to the colored region shown in Fig. 1(a).
In addition to the absorption cross-section, one can also reveal the contour plot for the supported extinction cross-section in TM or TE modes, i.e., $\sigma_n^{\text{ext} (a,b)}$ with a normalization factor $2\pi/(2n+1)\lambda^2$.
By demonstrating the extinction cross-section  with the power conservation intrinsically embedded in phase diagram,  one can clearly see that the region to support a zero extinction cross-section just corresponds to the invisible condition, i.e., $|a_n| = |b_n| = 0$, which illustrates another manifestation of the optical theorem.
This phase diagram not only reports the detail energy assignments among scattering and absorption powers, but also indicates the required phase and magnitude boundaries for all the  scattering coefficients.
Below, based on this phase diagram, we investigate the extreme limits on light scattering patterns.

To satisfy Kerker first and second conditions for extreme small magneto-dielectric spheres (as the dominant terms are the lowest-orders),  we need to satisfy the destructive or 
constructive  interferences between the TM and TE modes, i.e., $a_1 = -b_1$ or $a_1 = b_1$ for ZFS or ZBS, respectively.
Within quasi-static regime, the corresponding scattering coefficients for $a_1$ and $b_1$ can be expressed in terms of the permittivity $\epsilon$ and permeability $\mu$~\cite{opticaltheorem,phase}: 
\begin{eqnarray}
a_{1}&=&[1+i\frac{3\lambda^{3}}{2(2\pi a)^{3}}\frac{2+\epsilon}{\epsilon-1}]^{-1},\\
b_{1}&=&[1+i\frac{3\lambda^{3}}{2(2\pi a)^{3}}\frac{2+\mu}{\mu-1}]^{-1}.
\end{eqnarray}
Thus, the conditions $a_1 = -b_1$ or $a_1 = b_1$, lead to the famous formula  $\epsilon=(4-\mu)/(2\mu+1)$ or $\mu=\epsilon$ for ZFS or ZBS, respectively~\cite{note1}.
However, we wan to emphasize that as one can see in Fig. 1(a), there exists an additional implicit degree of freedom to satisfy the destructive or  constructive  interferences between the TM and TE modes. Now, to go one step further, we demonstrate how to utilize the phase diagram with this additional degree of freedom.

For the implicit parameters in the phase digram, we introduce two real numbers $\alpha_{n}^{(a,b)}$ and $\beta_{n}^{(a,b)}$ for $n$-th order spherical harmonic  wave in TM and  TE modes by re-writing the scattering coefficients  as 
\begin{eqnarray}
a_n &=& \frac{1}{1+i(\alpha^a_n+i\beta_n^a)},\\
b_n &=& \frac{1}{1+i(\alpha^b_n+i\beta_n^b)}.
\end{eqnarray}
In particular, the parameter $\beta_n^{(a,b)}$ accounts for the material lossy effects. Only when $\beta_n^{(a,b)} = 0$, the scattering system is made of lossless materials. 
For the allowed solutions in the phase diagram, the intrinsic power conservation automatically sets the range for these parameters, i.e.,  $\alpha_{n}^{(a,b)}=[-\infty,\infty]$ and $\beta_{n}^{(a,b)}=[0,-\infty]$, no matter what kind of geometric configurations or material compositions for scattering systems are chosen. 
Also, it should be noted that the border depicted by Blue-dashed curve in Fig. 1(a) reflects  a lossless boundary, i.e., $\beta_{n}^{(a,b)}= 0$.

To achieve ZFS,  we need  to have  $a_1 = -b_1$. It means that a phase difference $\pm \pi$ and the same magnitude of scattering coefficients are needed.
Nevertheless, as one can see from the phase diagram in Fig. 1(a), the only supported solutions are localized at ($\text{Arg}(a_n)= \pm \pi/2, |a_n|=0$) and ($\text{Arg}(b_n)= \mp \pi/2, |b_n|=0$).
In terms of  the implicit parameters, to approach ZFS,  $\alpha$ needs to go to  $\pm\infty$; while $\beta = 0$.
Only in these asymptotic cases,  the corresponding phase differences between $a_1$ and $b_1$ will be $\pm \pi$ while they have the same magnitudes.
The marked pair (II) corresponds to such a zero absorption and a zero scattering cross-sections, with the extinction cross-section $\sigma_n^{\text{ext}} = 0$.
In general, the scatterer can have the same magnitude in two lowest-order scattering coefficients, $\vert a_1\vert=\vert b_1\vert$, but the required phase difference $\pm \pi$ is impossible for non-invisible scattering system.
In other words, the ideal ZFS is impossible to be realized from a passive scatterer. Only nearly zero forward scattering can be approached.
Instead, to achieve  ZBS, one can easily find supported solutions in the phase diagram by looking for the pair of $a_{1} = b_{1}$.
Moreover,  a variety of solution pairs in the allowable region of phase diagram can be supported.

\begin{figure*}[ht]
\begin{center}
\includegraphics[width=14cm]{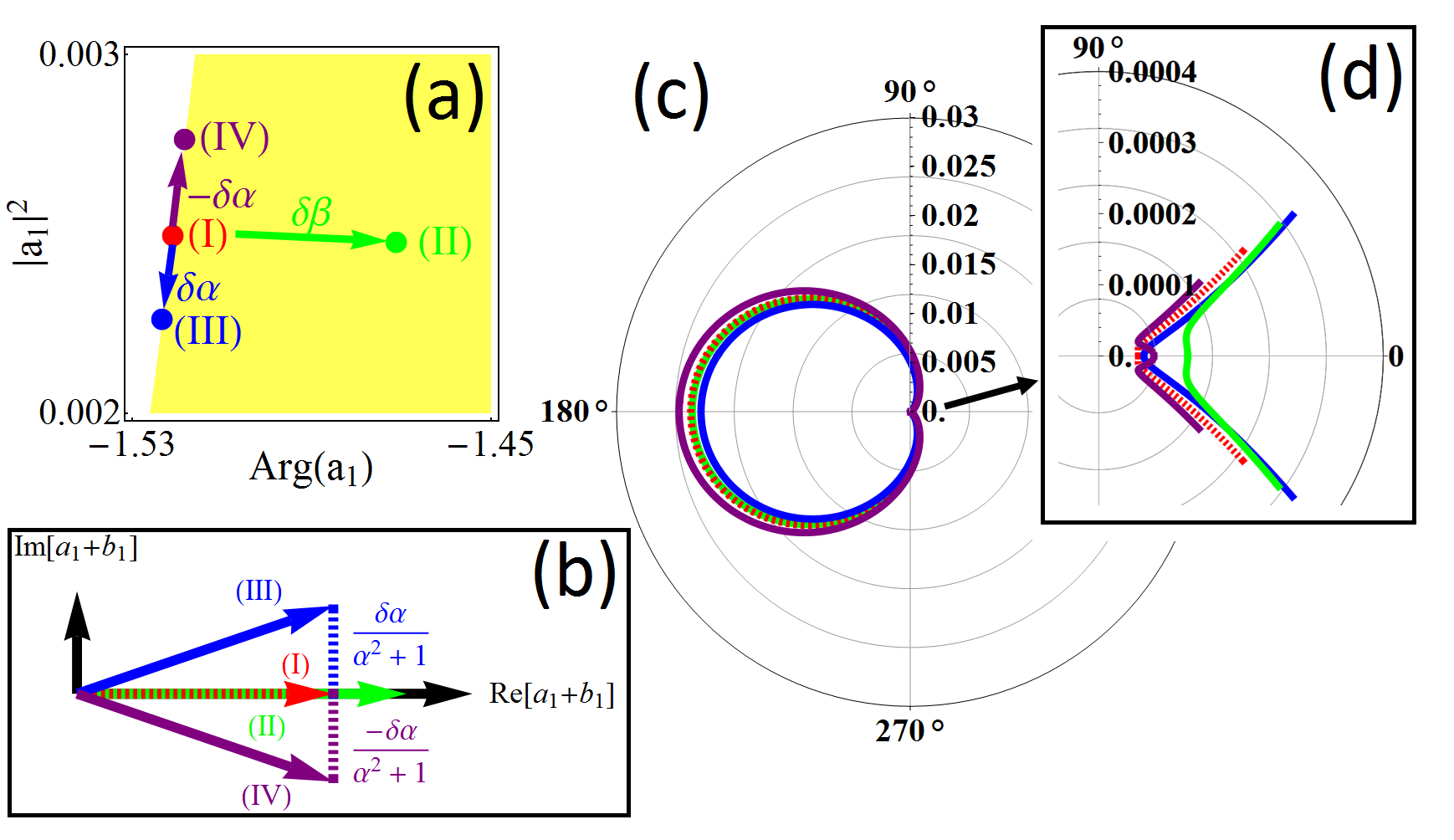}
\end{center}
\caption{(a) An unperturbed NZFS state marked as (I) in phase diagram is studied with the introduction of a small perturbation of $\delta \beta$ as marker (II), $\delta \alpha$ as marker (III), and $-\delta \alpha$ as marker (IV), respectively.
(b) The corresponding auxiliary complex space, defined by the Real and Imaginary parts of the scattering coefficient $a_1 + b_1$, for the unperturbed and perturbed states.
(c) The resulting 2D scattering patterns for the unperturbed and perturbed states given in (a), with the enlarged window in the forward direction shown in (d).} 
\end{figure*}

With the help of these implicit parameters,  one can generalize the Kerker second equation as
\begin{eqnarray}
\epsilon(\alpha)&=&\frac{3+\alpha(k_0 a)^3}{\alpha(k_0 a)^3-\frac{3}{2}},\\
\mu(\alpha)&=&\frac{-3+\alpha(k_0 a)^3}{\alpha(k_0 a)^3+\frac{3}{2}}.
\end{eqnarray}
Here, for the simplicity,  we have assumed that the scattering system is made of lossless materials, i.e.,  $\beta_{1}^{a}=\beta_{1}^{b}=0$, and also define $\alpha_{1}^{a}=-\alpha_{1}^{b} \equiv \alpha$ for the original condition $a_1 = -b_1$.
By eliminating the implicit parameter $\alpha$, one can come back to  the original second Kerker's condition, i.e., $\epsilon=(4-\mu)/(2\mu+1)$.
Based on Eqs. (8-9), one can expect that there exist a family of solutions to satisfy  Kerker second condition as the implicit parameter $\alpha$ varies accordingly.
In particular, perfect ZFS only happens when $\alpha = \pm \infty$, which is also the condition for invisible scatterers.
On the other hand, the inconsistency from the Kerker second condition to the optical theorem can be easily interpreted in phase diagram.
As for the exception solution in Kerker second condition, $\epsilon = \mu = -2$ ~\cite{break}, the corresponding implicit parameter  $\alpha=0$.
From the phase diagram, we can see that this exception solution gives ZBS, instead of ZFS.
In short, the improper results conducted from original Kerker second condition can be 
resolved from this set of generalized formulas given in Eqs. (8-9).

Even though a perfect ZFS condition is not allowed,  we can explore the possibility to minimize the forward scattering by defining the forward scattering efficiency:
\begin{eqnarray}
\eta^{fw}=\frac{\sigma^{fw}}{\sigma^{scat}},
\end{eqnarray}
as the ratio between forward scattering cross-section to the total scattering cross-section.
For the perfect ZFS, this forward scattering efficiency goes to zero; while a non-zero value gives a metric to quantify the forward scattering.
Then, for the lossless system, $\beta_n^{(a, b)} = 0$, the forward scattering efficiency has the following form:
\begin{eqnarray}
\eta^{fw}=\frac{3}{4(1+\alpha^{2})}.
\end{eqnarray}
Again, only when $\alpha\rightarrow \pm\infty$, we have the ZFS as $\eta^{fw} \rightarrow 0$.

To demonstrate the supported directional radiation patterns in phase diagram, in Fig. 1(b-g), we illustrate three extreme limits on light scattering: ZBS, NZFS (with a negligible extinction cross-section),  and the state with the same extinction cross-section as NZFS but including material lossy effect.
First, we consider the state with a maximum value in the extinction cross-section, i.e., $a_1 = a_2 = 1$, which is marked as (I) in Red-colors in the phase diagram of Fig. 1(a).
Both the electric dipole and magnetic dipole are at resonance, resulting in a giant scattering pattern shown in Fig. 1(b) and 1(c) for three-dimensional (3D) and  two-dimensional (2D) contour plots, respectively.
As $a_1 = b_1$ condition is satisfied, a clear ZBS pattern can be seen. 
By referring to the material properties through Eqs. (4-5), we have  $\epsilon=-2.005$ and $\mu=-2.005$.

Next, we choose the supported solutions closely to ZFS condition, i.e., the pair marked as (II) by Black-colors shown in Fig. 1(a).
Clearly, from the corresponding radiation patterns shown in Fig. 1(d-e), one can see an nearly ZFS pattern, but with an extremely low residue of scattering in the forward direction, with the corresponding extinction cross-section in the order of $10^{-6}$, see the Inset.
By Eqs. (4-5), the corresponding materials can be found as $\epsilon=-2.025$ and $\mu=-1.985$. 
As one can see from Fig. 1(a), there exist a family of supported solutions with the same value of extinction cross-section. 
For example,  the solution pair marked as (III) by Purple-colors in Fig. 1(a), reveals the scenario with the  same amount of extinction cross-section as that from the solution pair (II), i.e., localed in the same contour line,  but with the intrinsic material loss.
Clearly, the out-of-phase requirement is not satisfied for the ZFS. 
As shown in Fig. 1(f-g), even though the strengths of scattered electric field in the forward direction is the same as that  in the solution pair (II), i..e, note the scale now is 
in the order of $10^{-6}$, the corresponding radiation patterns in these two states are totally different.
The corresponding material parameters are $\epsilon=-2.33+0.63i$ and $\mu=-1.61+0.38i$ for the state marked as pair (III).
In terms of the forward scattering efficiency,  for these three parameter pairs: (I), (II), and (III) chosen in Fig. 1(a), the corresponding values are $\eta^{fw} = 3/4$, $0.0005$, and $0.47$, respectively.
Again, even though the last two pairs have the same extinction cross-sections, but there is a big difference in their forward scattering efficiencies.

With the help of the phase diagram, we can also investigate the robustness of supported NZFS states, by taking strength mismatch or material lossy effects into consideration.
In term of the explicit parameters, we introduce two small perturbation terms $\delta\alpha$ and $\delta\beta$ into the scattering coefficient $a_{1}$, i.e., $a_1=[1+i(\alpha+\delta\alpha+i\delta\beta)]^{-1}$.
In Fig. 2(a), we illustrate the perturbed solutions in the phase diagram, with the unperturbed one, marked as (I).
Specifically, a perturbation with $\delta\alpha= 0, \delta\beta = -1$, marked as (II),  moves the unperturbed state ($\alpha=20$) away from the lossless boarder, i.e., to the right-handed side; while  a  perturbation with $\delta\alpha= \pm 1, \delta\beta = 0$  moves the original state downward or upward along the lossless trajectory, i.e., to the markers (III) or (IV), respectively.

To illustrate the scattering patterns with small perturbations, we can expand the required ZFS condition, $a_1+b_1=2/(1+\alpha^2),$, to the first-order terms of $\delta \alpha$ and $\delta \beta$. That is
\begin{equation}
a_1+b_1\approx\frac{2-\delta\beta-i\delta\alpha}{1+\alpha^2},
\end{equation}
here $\delta\beta$ is a  negative value to account for material losses.
Then, in the auxiliary complex space for $a_1+b_1$, as shown in Fig. 2(b), one can clearly see that the amplitude of $a_1+b_1$ should be as small as possible in order to have an optimized NZFS.
When $\delta\alpha=0$ but  $\delta\beta\neq 0$, Eq. (12) can be reduced to  $a_1+b_1\approx(2-\delta\beta)(1+\alpha^2)^{-1}$, which represents the induced forward scattering due to material loss with an additional incremental amplitude $(-\delta\beta)(1+\alpha^2)^{-1}$.
On the other hand,  when  $\delta\beta=0$ but $\delta\alpha\neq 0$, Eq. (12) becomes 
$a_1+b_1\approx (2-i\delta\alpha)(1+\alpha^2)^{-1}$, which gives an extra contribution in the forward scattering by $(\pm\delta\alpha)(1+\alpha^2)^{-1}$.
With the increments in amplitude or phase from the perturbed solutions, we plot the the resulting scattering patterns in Fig. 2(c), and its enlarged ones in Fig. 2(d).
As one can see, the material lossy effect from a nonzero $\delta\beta$ significantly  deteriorates the NZFS. 
Nevertheless, with $\delta \alpha \neq 0$ but $\delta \beta = 0$, the NZFS states is more insensitive to a  mismatch in magnitudes only~\cite{note}. 
Again, these examples indicate that the NZFS state relies not only  on the perfect strength matching between $a_1$ and $b_1$, but  also more crucially on the phase difference to the out-of-phase condition.

\begin{figure}
\centering
\includegraphics[width=8.4cm]{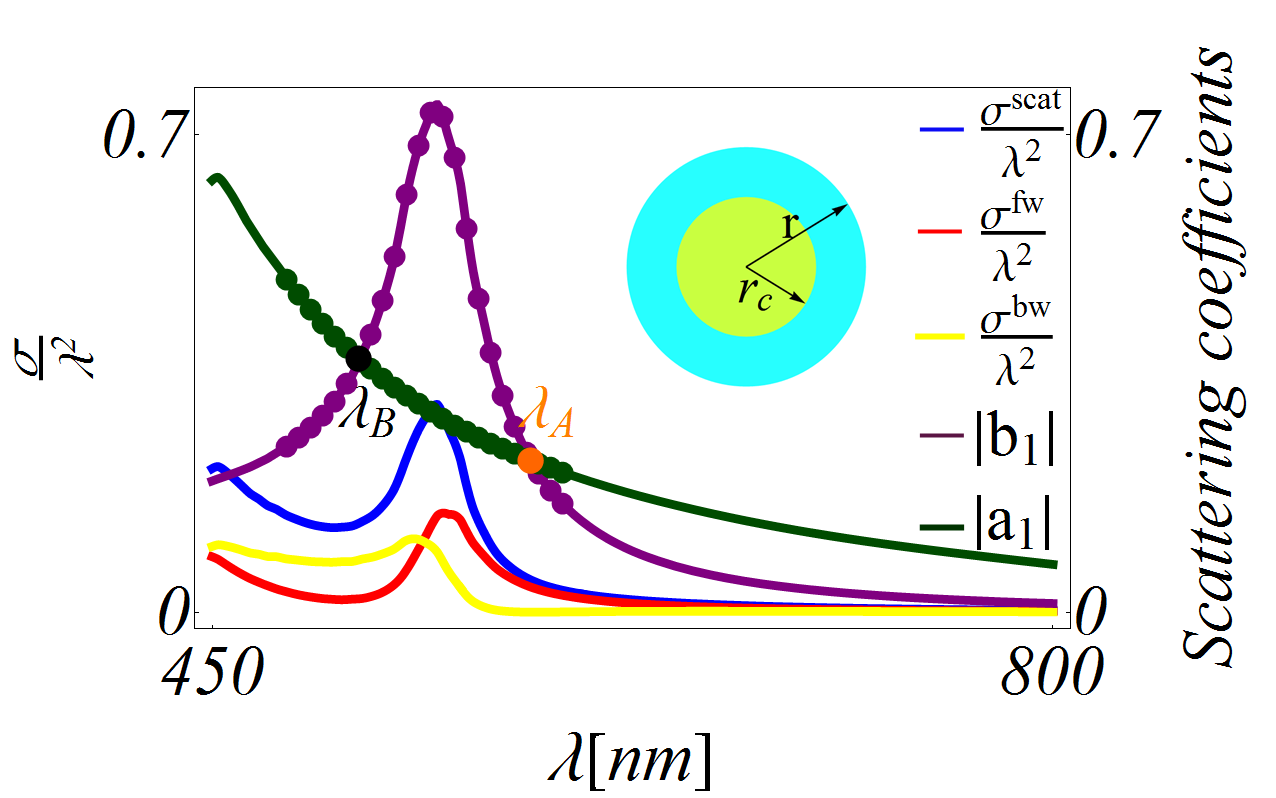}
\caption{The corresponding spectra for scattering cross-section (Blue-colored), forward scattering cross-section (Red-colored), , backward scattering cross-section (Yellow-colored)
and absolute values of two lowest-order scattering coefficients ($|a_{1}|$ and $|b_{1}|$) for a core-shell nanoparticle shown in the Inset. This core-shell particle consists of gold in core and silicon in shell, with the corresponding dispersion relation taken from the experimental data~\cite{palik}. } 
\end{figure}

In addition to the isotropic and homogeneous single-layered spherical scatterers, to realize directional scattering patterns such as NZFS or ZBS, we apply our phase diagram for core-shell particles, which recently are readily accessible with the experimental advance ~\cite{exp4}.
To realize such kind a core-shell configuration, one may use self-sacrificing template method to synthesize highly uniform nanoparticles with a tunable thickness ~\cite{photocatalytic, thickness}.
In particular,  we consider a core-shell nanoparticle with gold in core and silicon in shell,  as shown in the Inset of Fig. 3.
When realistic geometric size is taken for implementation, we set the outer radius $r$ to be $64$nm for the whole scatterer, and the core radius $r_c$ to be  $12$nm for the inside gold particle ~\cite{exp4}.
Moreover, the corresponding  material dispersion data is also taken into consideration from experimental data~\cite{palik}, for the optical wavelength region, i.e., $\lambda = 450$nm to $800$nm.

\begin{figure*}[ht]
\begin{center}
\includegraphics[width=18cm]{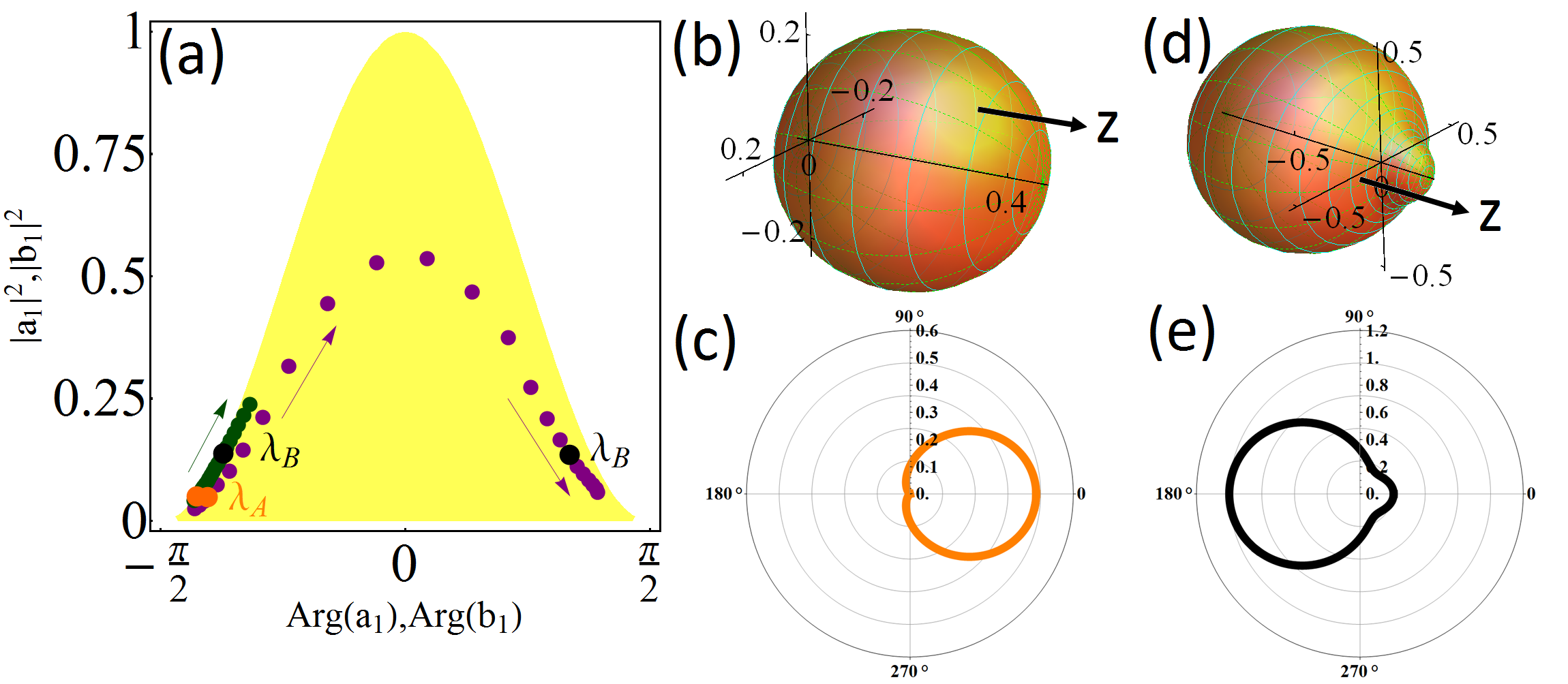}
\end{center}
\caption{Locations of the two lowest-order scattering coefficients, $a_{1}$ and $b_{1}$, at different wavelengths in the phase diagram, generated from the data points depicted in Fig. 3. Here, Green- and Purple-dots represent the  scattering coefficients for electric and magnetic dipoles, $a_1$ and $b_1$, respectively. The markers $\lambda_A$ and $\lambda_B$ highlight the two crossing-points marked in Fig. 3, i.e., for the ZBS and NZFS conditions, respectively.  The corresponding 3D-2D scattering patterns are plotted in (b-c) for the marker $\lambda_A$ for the ZBS condition; and  (d-e) for the markers $\lambda_B$ for the NFZS condition.} 
\end{figure*}

Based on the dispersion relations for gold and silicon, in Fig. 3, we also depict the corresponding  total, forward, and backward scattering cross-sections, $\sigma^{scat}$, $
\sigma^{fw}$, and $\sigma^{bw}$ (normalized to $\lambda^2$) by Blue-, Red-, and Yellow-colors, respectively.
Due to the normal mode resonance from TM mode, $b_1$, there exists a resonance peak around $550$nm.
In order to give an illustration on the scattering behavior, in the same plot, the magnitudes of two lowest-order scattering coefficients $|a_1|$ and $|b_1|$ are also depicted in Green- and Purple-colors, respectively.
One can see clearly that in the long wavelength limit, i.e., the incident wavelength $\lambda>600$nm, the electric dipole is always dominant, i.e., $|a_{1}| > |b_1|$ . 
However, near the resonance region, around $\lambda = 550$nm, the magnetic dipole can overwhelm the electric one, i.e., $|b_1| > |a_1|$.
When we approach the resonance condition by decreasing the wavelength, there exist two crossing points, marked as $\lambda_A = 583$nm and $\lambda_B = 511$nm in Fig. 3, having the same magnitude in the two scattering coefficients.

Even though with the help of the spectra in Fig. 3 we can expect to have exotic light scattering at the two crossing points, the underline physical picture to support ZBS or ZFS is not clear.
Instead,  to have a better understanding on the scattering properties for such a core-shell configuration, in Fig. 4(a), we plot all the trajectories at different wavelengths in phase diagram.
As the wavelength decreases (from long wavelength to short wavelength), i.e., indicated by the arrows, these two lowest-order scattering coefficients move in the supported region.
First of all,  in the long wavelength limit, both $a_1$ and $b_1$ are located at the same (left) side in the phase diagram, i.e., near the phase $-\pi/2$.
Then, as the wavelength decreases, they meet together at the crossing point $\lambda_A$, indicating the relative phase between them is $0$.
Now, we have $a_1 = b_1$. 
The resulting scatting patterns at this crossing point $\lambda_A$ is ZBS, as clearly demonstrated in Fig. 4(b) and 4(c) for 3D and 2D plots, respectively. 
%{\color{red} The corresponding forward scattering efficiency $\eta^{fw}$ is $0.78$.}

On the other hand, due to the magnetic dipole resonance, the trajectory for the scattering coefficient $b_1$ should pass through the center ($\text{Arg}(b_1) = 0$) in phase diagram.
However, for the scattering coefficient $a_1$, it would stay in the same (left) side as long as the electric dipole resonance is not excited.
Then, we have the possibility to generate the second crossing point at $\lambda_B$, at which  the scattering coefficients $a_1$ and $b_1$ form a complex conjugated pair in the phase diagram.
As we discussed above, now, we have $|a_1| =|b_1|$, but the phase difference is not large enough to meet the out-of-phase shift. 
The resulting radiation pattern can be expected to be  a NZFS only, as shown in Fig 3(d-e).
In this practical example, we have the forward scattering efficiency $\eta^{fw}=0.15$.

With the core-shell nanoparticle illustrated above, we demonstrate a direct interpretation of directional scattering patterns through the  supported trajectories in phase diagram. 
In particular, by varying the incident light wavelength, we predict to have a significant change in the directional scattering patterns, i.e., from  ZBS to NZFS with a single configuration.
Although our discussion is only limited to spherical structures, the concept and approach here with  phase diagram can be easily applied to non-spherical geometry.
In addition to the lowest-orders, one can also  take higher-order spherical harmonic channels into consideration as a general extension.

In summary, with the supported solutions in the phase diagram, we revisit Kerker first and second conditions on the zero backward scattering (ZBS) and zero forward scatting (ZFS).
In addition to the explanations on the scattering coefficients, we give a clear physical picture on the physical bounds  on scattering distributions.
The known problems with the  inconsistency and related exception solution in the original Kerker second condition can be resolved in phase diagram.
We also reveal that there exist a set of implicit parameters ($\alpha$ and $\beta$) to compose the scattering coefficients, and derive a generalized Kerker condition.
In the phase diagram, a perfect ZFS requires the asymptotic conditions as the implicit parameter $\alpha$ goes to $\pm \infty$.
To be consistent with the optical theorem, only nearly zero forward scattering (NZFS) can be realized for passive electromagnetic scatterers with a finite value of the implicit parameter $\alpha$. 
The robustness of NZFS is also investigated in phase diagram with a small variation in  strength or material losses on directional scattering patterns.
To implement the NZFS and ZBS, a  core-shell nanoparticle  is proposed, with the real material dispersion relation taken into consideration.
Through the supported trajectories in phase diagram, we predict a change from ZBS to NZFS in the scattering patterns with the same geometric configuration, but just decreasing the incident wavelength in the optical domain.
With the advances in nanoparticle syntheses,  these extreme limits on light scattering provided by phase diagram can be readily  to promote the  designs on directional scattering nano-devices.

\section*{Acknowledgments}
This work is supported in part by the Ministry of Science and Technology, Taiwan, under the contract No. 105-2628-M-007-003-MY4, and by the Australian Research Council.

\bibliography{achemso-demo}

\begin{thebibliography}{99}
\bibitem{note1} In extremely  subwavelength case $\lambda^3/a^3\gg 1$, Milton KerKer neglected the real part of the denominator in  Eq.(4) and Eq.(5) in calculation, as a result, he obtained the famous Kerker second condition.
However, such approximation would violate power conservation, as pointed out in the phase diagram.
Ref.~\cite{opticaltheorem} also discussed this issue.
\bibitem{note} It should note that small variation of $\delta\alpha$ would also change a little phase compared to $b_1$ clearly seen in phase diagram.
But this change is quite smaller than the magnitude mismatch effect.
\end{thebibliography}

\end{document}